\documentclass[12pt, letterpaper, oneside]{article}
\pdfoutput=1

\usepackage{graphicx}
\usepackage[body={6.5in, 9in}, right=1in, top=1in]{geometry}
\usepackage{amssymb}
\usepackage{amsmath} 
\usepackage{color}
\usepackage{hyperref}

\newcommand\ee{\end{equation}}
\newcommand\be{\begin{equation}}
\newcommand\eea{\end{eqnarray}}
\newcommand\bea{\begin{eqnarray}}

\newcommand\di{\partial}

\newcommand{\la}{\label}
\newcommand{\eq}[1]{Eq.~(\ref{#1})}

\newcommand{\gre}{\color{green}}

\def\bea{\begin{eqnarray}}
\def\eea{\end{eqnarray}}	
\def\ba{\begin{array}}
\def\ea{\end{array}}
\def\l{\left}
\def\r{\right}
\def\d{\partial}
\def\fr{\frac}
\def\f{\frac}
\def\dd{{\rm d}}

\def\a{\alpha}

\def\m{\mu}
\def\n{\nu}

\def\bi{\begin{itemize}}
\def\ei{\end{itemize}}

\def\nn{\nonumber}

\begin{document}

\begin{center}
\LARGE{\textbf{(Re-)Inventing the Relativistic Wheel: \\
Gravity, Cosets, and Spinning Objects}} \\[1cm]
\large{Luca V. Delacr\'etaz$^{\rm a \rm c }$, Solomon Endlich$^{\rm b}$, Alexander Monin$^{\rm b}$, Riccardo Penco$^{\rm a}$ and Francesco Riva$^{\rm b}$}
\\[0.4cm]

\vspace{.2cm}
\small{\textit{$^{\rm a}$ Department of Physics and ISCAP, \\ 
Columbia University, New York, NY 10027, USA}}

\vspace{.2cm}
\small{\textit{$^{\rm b}$ Institut de Th\'eorie des Ph\'enom\`enes Physiques, EPFL,\\
1015 Lausanne, Switzerland}}

\vspace{.2cm}
\small{\textit{$^{\rm c}$ Institute for Advanced Study,  Hong Kong University \\ of Science and Technology,
Hong Kong}}

\end{center}

\vspace{.2cm}

\begin{abstract}

Space-time symmetries are a crucial ingredient of any theoretical model in physics. Unlike internal symmetries, which may or may not be gauged and/or spontaneously broken, space-time symmetries do not admit any ambiguity: they {\em are} gauged by gravity, and any conceivable physical system (other than the vacuum) is bound to break at least some of them. Motivated by this observation, we study how to couple gravity with the Goldstone fields that non-linearly realize spontaneously broken space-time symmetries. This can be done in complete generality by weakly gauging the Poincar\'e symmetry group in the context of the coset construction. To illustrate the power of this method, we consider three kinds of physical systems coupled to gravity: superfluids, relativistic membranes embedded in a higher dimensional space, and rotating point-like objects. This last system is of particular importance as it can be used to model spinning astrophysical objects like neutron stars and black holes. Our approach provides a systematic and unambiguous parametrization of the degrees of freedom of these systems.

\end{abstract}

\newpage





\section{Introduction}

Symmetries are arguably one of the most important properties of physical systems. In particular, when spontaneously broken, they lead to model-independent predictions for the low-energy spectrum of excitations and their interactions. In relativistic theories with spontaneously broken internal symmetries, Goldstone's theorem ensures the existence of one gapless mode -- a Goldstone boson -- for each broken symmetry generator. Moreover, spontaneously broken symmetries are still symmetries of the effective action (barring anomalies), and they constrain the dynamics in a subtle way by acting non-linearly on the Goldstone fields. A systematic method to build an effective action for the Goldstone modes that is invariant under the non-linearly realized symmetries was developed by Callan, Coleman, Wess and Zumino in the late sixties and it is known as the \emph{coset construction}~\cite{Coleman:1969sm,Callan:1969sn}. The virtue of this method is that it relies solely on symmetry considerations and it allows one to be  agnostic about the symmetry breaking mechanism, which in general could be very complicated and even strongly coupled. The coset construction was later extended to the breaking of \emph{space-time} symmetries~\cite{Volkov:1973vd,ogievetsky:1974ab}, in which case several interesting subtleties arise. It is well known, for instance, that in this case the number of Goldstone modes does not need to equal that of broken symmetries~\cite{Ivanov:1975zq,Low:2001bw}. Moreover, Goldstone excitations do not need to be massless~\cite{Nicolis:2012vf,Kapustin:2012cr,Watanabe:2013uya,Nicolis:2013sga} or stable, the UV completion in these systems may occur in non-standard ways~\cite{Endlich:2013vfa} and even the issue of superluminality becomes subtle~\cite{deRham:2013hsa,Creminelli:2013fxa,Creminelli:2014zxa,deRham:2014lqa}.

The spontaneous breaking of space-time symmetries is an interesting phenomenon not only because of the aforementioned subtleties, but also because of its ubiquity:  any state of matter which is not the vacuum must break at least some space-time symmetries. For instance, even a state with a single point particle at rest breaks boosts by selecting a preferred reference frame. On the contrary, while it is certainly possible to consider states that spontaneously break any given internal symmetry, one is not \emph{forced} to do so: for example, if the above point particle is charged under a $U(1)$ symmetry, the corresponding state is an eigenstate of the charge and does not break $U(1)$. Internal and space-time symmetries also differ in another respect: while Nature has provided us with both global and gauged internal symmetries, there is no ambiguity when it comes to space-time symmetries - they \emph{are} gauged by gravity. The overarching goal of this paper is to illustrate how an appropriate extension of the coset construction can be used to describe the coupling between gravity and those systems whose ground state breaks some space-time symmetries.

In an effort to make the paper as self-contained as possible, we review the coset construction for both internal and space-time symmetries in Section \ref{coset construction}. 
Then in Section \ref{gravity_and_coset}, we show how the coset construction can be fruitfully employed to formulate ordinary General Relativity from an ``algebraic'' point of view.\footnote{A similar approach was used in~\cite{ArmendarizPicon:2010mz} to study modified theory of gravity with spontaneous breaking of local Lorentz symmetry.} After that we move on to consider a few instructive examples of systems with broken space-time symmetries. In Section \ref{superfluid} we show how to use the coset construction to couple relativistic superfluids to gravity.\footnote{For a recent application of coset techniques to relativistic superfluids in the absence of gravity, see~\cite{PhysRevD.89.045002}.} Such systems have been studied extensively in the context of cosmology as a possible mechanism to generate inflation~\cite{ArkaniHamed:2003uz} or to modify the large distance behavior of gravitational interactions~\cite{ArkaniHamed:2003uy}. From our perspective, the interesting aspect about superfluids is that they are possibly the simplest system in which a combination of space-time and internal symmetries is broken down to a diagonal subgroup. 

The second example that we will consider is that of relativistic membranes (Section \ref{membranes}). These objects have attracted a great amount of theoretical interest in the last few decades and have appeared in a variety of different contexts: as D-branes~\cite{Johnson:2000ch} in string theory or  as domain walls~\cite{Vachaspati:2006zz}, in models of extra-dimensions~\cite{Rattazzi:2003ea,Csaki:2004ay} and in connection to Galileon theories~\cite{deRham:2010eu,Goon:2011uw}, etc. 
We will show how gravity in the bulk can be coupled to Goldstones bosons that propagate on a lower-dimensional submanifold, and how the geometric language usually employed to describe membranes arises naturally from the coset construction. In the case where the lower-dimensional submanifold is one-dimensional and its only tangent vector is time-like, one easily recovers the correct action for a relativistic point-particle (Section \ref{point particles}).

Finally we turn to the study of  relativistic spinning particles coupled to gravity (Section~\ref{spinning particle}). These are objects of great  astrophysical importance as they provide a long-wavelength effective description of compact objects, such as spinning black holes  and neutron stars  and constitutes the most important application of these techniques. To the best of our knowledge, relativistic spinning objects in a gravitational field were first studied using effective field theory techniques in Refs.~\cite{Porto, Porto:2006bt} (and later improved upon in Ref.~\cite{Porto:2008tb}), by extending the results of Hanson and Regge's~\cite{Hanson:1974qy} to a curved space-time.\footnote{See also Ref.~\cite{Balachandran:1979ha} for a different approach to relativistic spinning particles in gravitational fields (with torsion).} In this approach, some covariant constraints are imposed on the low-energy effective Lagrangian in order to eliminate spurious degrees of freedom~\cite{Hanson:1974qy}. These constraints have a clear physical interpretation, but become cumbersome to implement at higher order in the derivative expansion. It is here where the usefulness of the coset construction becomes most clear. 

In our approach, all unphysical degrees of freedom are systematically removed by imposing the appropriate inverse Higgs constraints, which can be solved exactly and whose simple solutions are valid to all orders. After that, it becomes straightforward to systematically write down a relativistic effective action for spinning particles coupled to gravity. The derivative expansion is controlled by the ratio $\omega_R /\omega_N$ between the frequency of rotation and the typical normal mode frequency of the rigid body. For completeness, a quick summary of our notation and conventions as well as some technical details have been included in the appendices.



\section{The coset construction} \la{coset construction}

We start with a brief review of the coset construction, which can be safely omitted by the reader already familiar with this formalism. For later convenience, we will adopt a notation that applies to the breaking of internal~\cite{Coleman:1969sm,Callan:1969sn} and space-time symmetries~\cite{Volkov:1973vd, ogievetsky:1974ab} alike.  For a thorough discussion of the coset construction for internal symmetries only, we refer the reader to~\cite{Weinberg:1996kr}.

The coset construction provides a systematic method of writing down an effective action for Goldstone bosons using the pattern of symmetry breaking as the only input. For definiteness, let us therefore consider a symmetry group $G$ (which includes Poincar\'e as a subgroup) and assume that the ground state spontaneously breaks it down to a subgroup $H$. We can then subdivide the generators of $G$ into three groups:
\bea
X_\alpha &=& \mbox{broken generators} \nonumber \\
\bar{P}_a &=& \mbox{unbroken translations} \nonumber \\
T_A &=& \mbox{other unbroken generators}. \nonumber 
\eea
Notice that both the $X$'s and the $T$'s will in general contain some space-time and some internal generators. The effective action for the Goldstone bosons must be invariant under the whole symmetry group $G$. However, the broken symmetries generated by the $X_\alpha$'s and the unbroken translations $\bar P_a$'s will be realized nonlinearly on the Goldstone fields.  Hence, it is a non-trivial task to write down all possible $G$-invariant combinations of the Goldstones and derivatives. The coset construction is designed precisely to address this problem.

The starting point of the coset construction is a local parametrization of the coset $G/H_0$, where $H_0$ is the subgroup of $H$ generated by the $T_A$'s. This parametrization reads\footnote{Notice that the pre-factor $e^{i y^a (x) \bar{P}_a}$ on the RHS of equation (\ref{cosetrep}) is usually omitted when considering only \emph{internal} broken symmetries~\cite{Weinberg:1996kr}, because in that case it does not play any role. It becomes however important when dealing with broken \emph{space-time} symmetries. }
\be \la{cosetrep}
\Omega (y, \pi) \equiv e^{i y^a (x) \bar{P}_a} e^{i \pi^\alpha (x) X_\alpha},
\ee
and can be thought of as the most general group element generated by the $X_\alpha$'s and the $\bar P_a$'s using coordinate-dependent parameters. The transformation properties of the Goldstones under a generic element $g$ of the symmetry group $G$ can be derived from the relation~\cite{ogievetsky:1974ab} 
\be \la{transrule}
g \, \Omega (y, \pi) = \Omega (y', \pi') \, h(y, \pi, g),
\ee
where $h(y, \pi, g)$ is a Goldstone- and coordinate-dependent element of the unbroken subgroup~$H_0$ that guarantees that the \emph{form} of \eq{cosetrep} is preserved by the $g$ transformation. The unbroken element $h(y, \pi, g)$ can be calculated for any given $g$ using only the algebra of the group $G$, but for our practical purposes this will not be necessary.  Equation (\ref{transrule}) then \emph{defines} the transformation rules for the $y$'s and the $\pi$'s. In particular, the  Goldstones $\pi$ will usually transform nonlinearly, while  the $y$'s will transform like cartesian coordinates under unbroken Poincar\'e transformations.\footnote{For instance, the simplest case one can consider is the one in which $g$ is simply an unbroken translation, i.e. $g =e^{i \epsilon^a \bar P_a}$. In this case, it is easy to see that the $y$'s transform like Cartesian coordinates under translations, i.e. $y' (x)= y (x)+ \epsilon$, while the $\pi$'s do not transform, i.e.  $\pi' (x) = \pi (x)$, and $h(y, \pi, g) =1$. These transformations rules are particularly simple due to the fact that we included the unbroken translations in the coset parametrization.}
In fact, on a Minkowski background it is always possible to choose the $x$ coordinates in such a way that $y^a (x)\equiv x^a$ everywhere.   However, on a curved background this will not always be possible, and in this case the $y$'s should be thought of as locally inertial coordinates at some point within the patch described by the $x$ coordinates (see Appendix \ref{gravity_and_coset:app} for more details).

Starting from the coset parametrization (\ref{cosetrep}), we can define the Maurer-Cartan form $\Omega^{-1} d \Omega$. Its components can be calculated explicitly using only the commutation relations among the various generators, and the result can be expressed as a linear combination of all the generators:
\be \la{MCform}
\Omega^{-1} \d_\mu \Omega =  E_\mu{}^a  (\bar P_a + \nabla_a \pi^\alpha X_\alpha + A_a^B T_B ).
\ee
One can derive the transformation properties of the coefficients that appear in the above linear combination starting from the transformation rule (\ref{transrule}). In particular, it can be shown~\cite{ogievetsky:1974ab} that the coefficients $E_\mu{}^a$ play the role of a vierbein, in the sense that the integration measure $d^d x \det E$ transforms like a scalar under all the symmetries and is covariant under an arbitrary change of the $x$ coordinates. In other words, the factor $\det E$ ensures that the coset construction can be carried out using an arbitrary coordinate system (i.e. not necessarily Cartesian). Notice that the vierbein $E_\mu{}^a$ becomes trivial when all the $X_\alpha$'s are generators of internal symmetries and the $x$'s are Cartesian coordinates.

The quantities $\nabla_a \pi^\alpha$ can be thought of as covariant derivatives for the Goldstone fields, since they transform covariantly under \emph{all} symmetries:
\be
\nabla_a \pi^\alpha (x) \quad \stackrel{g}{\longrightarrow} \quad \nabla_a \pi'^{\alpha } (x) = h_a{}^b (y,\pi,g) h_\beta{}^\alpha (y,\pi,g) \nabla_b \pi^\beta (x) \, ,
\ee
where the $h_a{}^b$ and $h_\beta{}^\alpha$ matrices are some representations of the group element $h(y, \pi, g) \in~H_0$. Thus, the covariant derivatives $\nabla_a \pi^\alpha$ transform according to a field- and coordinate-dependent representation of the unbroken subgroup $H_0$. 

Finally, it can be shown that the coefficients $A_a^B$ transform like a connection~\cite{ogievetsky:1974ab}, and can be used to define higher covariant derivatives of the Goldstone fields:
\be \la{covdev}
\nabla_a^H  \equiv [(E^{-1})_a{}^\mu \d_\mu + i A_a^B T_B ] \ .
\ee
These covariant derivative can also act on additional matter fields that transform in some linear representation of the unbroken group $H_0$. 

One can then build the most general Lagrangian density that is invariant under the full symmetry group $G$ and independent of the particular choice of coordinates $x$ simply by taking contractions of all the possible covariant derivatives (e.g. $\nabla_a \pi^\alpha, \nabla_a^H \nabla_b \pi^\alpha, ...$ ) that are manifestly invariant under the unbroken  subgroup $H_0$.

\subsubsection*{Gauge symmetries}

The coset construction reviewed above can be appropriately modified to describe gauge symmetries as well. If a subgroup $G' \subseteq G$ with generators $V_I$ is gauged, then one must simply replace the partial derivative in the definition of the Maurer-Cartan form with a covariant derivative, i.e.
\be
\label{how_to_introduce_gauge}
\Omega^{-1} \d_\mu \Omega  \quad \to \quad \Omega^{-1} D_\mu \Omega \equiv \Omega^{-1} (\d_\mu + i \tilde{A}_\mu^{I} V_I ) \Omega.
\ee
This modified Maurer-Cartan form can  be decomposed like on the RHS of equation (\ref{MCform}), and the vierbein $E_\mu{}^a$, the covariant derivatives $\nabla_a \pi^\alpha$ and the connection $A_\mu^B$ will now depend also on the gauge fields $\tilde{A}_\mu^I$. It is easy to see that $\Omega^{-1} D_\mu \Omega$ is indeed invariant under a local transformation
\be
\Omega \to g(x) \Omega, \qquad \quad \tilde A _ \mu\to g (x)\tilde A _ \mu g ^ {-1}(x)  - i g(x)  \, \partial _ \mu g ^ {-1}(x) \qquad \mbox{with  } g(x) \in G'.
\ee
If the gauged generators $V_I$ contain some of the broken generators $X_\alpha$, then one can make a gauge transformation and set to zero some of the Goldstones $\pi^\alpha$: this amounts to working in the unitary gauge. In this paper, we will use the procedure defined by (\ref{how_to_introduce_gauge}) to introduce gravitational interactions by gauging the Poincar\'e group.

\subsubsection*{Inverse Higgs constraints}

In the coset parametrization (\ref{cosetrep}), we have assigned one Goldstone field to each broken symmetry generator $X_\alpha$. However, it is well known that whenever space-time symmetries are broken, the counting of Goldstone modes becomes subtle and the usual Goldstone theorem does not apply~\cite{Low:2001bw}. Within the context of the coset construction, the possible mismatch between the number of broken generators and that of Goldstone modes follows from what is known as the \emph{inverse Higgs mechanism}~\cite{Ivanov:1975zq}. Such phenomenon can be summarized as follows: whenever
\begin{description}
\item[\emph{i.}]  the commutators between an unbroken translation $\bar P$ and a broken generator $X$ contains another broken generator $X'$, i.e. $[\bar P, X' ] \supset X$, and
\item[\emph{ii.}] $X$ and $X'$ do not belong to the same multiplet under $H_0$, 
\end{description}
one can set to zero the covariant derivative of the Goldstone of $X$ in the direction of $\bar P$ (denoted as $\bar \nabla \pi$) and solve such a condition---which is known as an \emph{inverse Higgs constraint}---to eliminate the Goldstones of $X'$ from the low energy spectrum of excitations in a way which is compatible with all the symmetries. Since this kind of constraints will play a crucial role for the systems that will be discussed in this paper, we conclude this section by briefly discussing their physical origin. 

From a UV perspective, there are two complementary ways of understanding the inverse Higgs phenomenon. When provided with an explicit order parameter, it can be the case that the Goldstones associated with the broken generators do not describe independent degrees of freedom. That is, there is a non-trivial solution to the equation
\be
\left(\pi X+\pi' X'\right) \langle \Phi \rangle=0 \, ,
\ee
where $\langle \Phi \rangle$ is the expectation value of the order parameter~\cite{Low:2001bw}. From this perspective, imposing an inverse Higgs constraint is equivalent to ``fixing a gauge'' in order to eliminate redundant degrees of freedom~\cite{Nicolis:2013sga}. However, this is not always the case as there are symmetry breaking patters where it is consistent to impose such a constraint but no such redundancy can exist \cite{Endlich:2013vfa}. This leads us to a second possiblity: if there is no such overcounting of degrees of freedom, one can instead view the inverse Higgs constraints as arising dynamically in the low energy limit. Indeed,  conditions \emph{i.} and \emph{ii.} guarantee that the covariant derivative of the Goldstone of $X$  contains a term linear in $\pi'$ and with no derivatives, which means that a generic action will contain a ``mass term'' for $\pi'$, i.e. $\pi'$ is gapped. Hence, at energies below this gap we can integrate out $\pi'$ to obtain an effective action for the remaining Goldstones. In this picture, the inverse Higgs constraint can be interpreted as coming from the equation of motion for $\pi'$. In general, such equation of motion will not be simply $\bar \nabla \pi =0$, but rather it will be equivalent to setting to zero a generic linear combination of $\bar \nabla \pi$, other covariant derivatives that transform in the same representation as $\bar \nabla \pi$~\cite{Ivanov:1975zq} as well as higher order covariant derivatives. From the gauge fixing perspective this would seem like an overly complicated gauge fixing condition, but from the gapped Goldstone perspective it is clearly the most natural constraint to impose. The important point, though, is that the effective theory does not depend on the exact form of the inverse Higgs constraints: once the derivative expansion is correctly implemented, the difference between imposing a ``generalized inverse Higgs constraint'' or the simplest possible one corresponds only to a redefinition of the coupling constants in the effective Lagrangian. Hence, throughout this paper we will always impose the simplest possible inverse Higgs constraints without any loss of generality.



\section{General Relativity from a Coset Perspective}\label{gravity_and_coset}

Before introducing gravity in theories with spontaneously broken space-time symmetries, it is instructive to review how ordinary General Relativity (GR) can be derived from a coset construction where the Poincar\'{e} group $ISO(3,1)$ is gauged\footnote{Other backgrounds, such as those with a large cosmological constant in comparison with the typical energies we are interested in, can be studied using (anti-) deSitter group as starting point~\cite{Ivanov:1981wn, Ivanov:1981wm, Wise:2009fu}.}  and translations are non-linearly realized~\cite{Ivanov:1981wn}.\footnote{Notice that our approach differs form that of~\cite{Borisov:1974bn,Ivanov:1976pg}  and, more recently,~\cite{Goon:2014ika} which treat the gauge fields as Goldstone bosons associated with the breaking of local symmetries down to global ones.} Therefore, the coset construction we will carry out will be based on the coset $ISO(3,1)/SO(3,1)$, which can be conveniently parametrized as
\begin{equation}\label{cosetgravity}
\Omega\equiv e^{i y ^ a (x) P_a} \; .
\end{equation}
Notice, that the action of diffemorphisms amounts to relabeling  the space-time coordinates $x ^ \mu$, which do not transform under translations (for more details, see Appendix~\ref{gravity_and_coset:app}). Following the discussion in the previous section, we now introduce the Maurer-Cartan form associated with the coset parametrization (\ref{cosetgravity}) and we introduce gauge fields for translations ($\tilde e_\mu ^ a$) and for Lorentz transformations ($\omega_\mu^{ab}$). The Maurer-Cartan form then reads
\begin{equation}\label{SuperMaurerCartan}
\Omega^ {-1} {D}_\mu \Omega  \equiv e^{-i y ^ a (x) P_a} 
\left ( \partial _ \m +  i \tilde e_\mu{}^ a P _ a + \f {i} {2} \omega_\mu^{ab} J _ {ab}\right ) e^{i y ^ a (x) P_a}
= i e_\mu{}^a P_a + \frac{i}{2} \omega_\mu^{ab} J_{ab}\,,
\end{equation}
where in the last step we defined
\bea \label{e}
e _ \mu{}^ a = \tilde e _ \mu{}^ a + \partial _ \mu y ^ a +  \omega ^{ab} _ {\mu} \, y_b \,.
\eea
%
According to the previous section, the fields $e_\mu{}^a$ should now be regarded as a vierbein. In particular, they can be used to build an invariant volume element $d^d x \det e$. As a matter of fact, we will see that $e_\mu{}^a$ is indeed the usual vierbein that appears in the tetrad formalism~\cite{weinberg:1972bo}, in the sense that it defines the metric via $g_{\mu\nu} \equiv \eta_{ab} e _ \mu{}^ a e _ \nu{}^ b$.

Following the standard coset ``recipe'' we can now use the coefficients that appear in front of the unbroken Lorentz generators in (\ref{SuperMaurerCartan}) to define the covariant derivative of matter fields that transform linearly under Lorentz transformations:
\be
\nabla ^ L _ a \equiv (e^{-1})_a{}^\mu ( \partial _ \m + \f {i} {2} \omega _\m ^ {b c} J _ {b c} ) \, .
\label{cov_der_L}
\ee
By now, it should be clear that $\omega_\mu^{ab}$ is nothing but the spin connection that is usually introduced in the tetrad formalism~\cite{weinberg:1972bo}.
The vierbein (\ref{e}) and the covariant derivative (\ref{cov_der_L}) are the only necessary ingredients to describe the non-linear realization of translations and the local action of the Poincar\'e group. Then, neglecting for simplicity additional matter fields, the most general action that is Poincar\'e and diffeomorphism invariant takes the simple form
\be\la{SGR}
S=\int d^4 x \det e \,  \mathcal{L}(\nabla ^ L _a)\, ,
\ee
where it is understood that the indices of the covariant derivatives $\nabla ^ L _a$ must be contracted in a Lorentz-invariant fashion. From here on, one can identify the usual curvature invariants by proceeding as usual. In particular, one can use the fact that the commutator of two covariant derivatives acting on, say, a vector field gives
\be
[ \nabla ^ L _ a, \nabla ^ L _ b ] V^c = R^c{}_{dab}V^d - T_{ab}{}^d \nabla ^ L _ d V^c,
\ee
where $R^c{}_{dab}$ and $T_{ab}{}^d$ are the components of the Riemann and torsion tensor respectively w.r.t. the orthonormal frame defined by the vierbein $e_\mu{}^a$ (see Appendix~\ref{gravity_and_coset:app} for more details). Then, at lowest order in the derivative expansion the effective action (\ref{SGR}) reduces~to 
\be
S =\f {1} {16 \pi G} \int \det ( e ) d ^ 4 x \, \left [ R^ {a b}{}_ {ab} 
+  c_1 \, T_{ab}{}^ c T^{ab}{}_c +  c_2 \, T_{abc} T^{acb} +  c_3 \, T_{ab}{}^b T^{ac}{}_c + \cdots \right ] \, ,
\label{EH_gravity0}
\ee
where $e ^ \m{}_ a$ is defined as the inverse of $e _ \m{}^ a$ and, comparison with experiments would reveal that $G$ corresponds to Newton constant; $c_1$, $c_2$, and $c_3$ denote dimensionless coefficients and the dots stand for higher-order terms in the derivative expansion.

The action \eq{EH_gravity0} has more degrees of freedom than those associated with standard GR. However, the equations of motion for $\omega^ {a b} _ \m$ to lowest order in derivatives, are
\be\label{omegaofe0}
\omega ^ {a b} _ \m(e) = \f {1} {2} \l [ e ^ {\n a} (\partial _ \m e _ \n{} ^ b - \partial _ \n e_ \m{} ^ b ) +e _ {\m c} e^ {\n a} e ^ {\lambda b} 
\partial _ \lambda e _ \n{} ^ c - (a \leftrightarrow b) \r ] \, ,
\ee
and give a non-dynamical condition on $\omega^ {a b} _ \m$ that is precisely the standard relation between a tetrad and a spin connection for gravity in the vierbein formalism~\cite{weinberg:1972bo}.\footnote{Notice that even in the presence of additional matter fields, the equation of motion for the spin connection can still be solved algebraically at lowest order in the derivative expansion. The solution will in general differ from the one in equation (\ref{omegaofe0}), but upon plugging the new solution into the effective action, one obtains a torsion-free theory with shifted coefficients in the matter effective action. Therefore, in our context, treating the spin connection $\omega$ as an independent variable is equivalent to imposting the torsion free condition.} With this, the action (\ref{EH_gravity0}) reduces to the famous Einstein-Hilbert action (at lowest order in derivatives).  Alternatively, the condition \eq{omegaofe0} can be derived directly by noticing that it is consistent---for it transforms in a covariant way---to enforce the torsion tensor to zero, i.e. $T _ {ab}{}^ c = 0$ \, (somewhat in analogy with the inverse Higgs procedure discussed in the previous section).\footnote{One could alternatively choose to set  the curvature tensor to zero, in which case one obtains a teleparallel theory of 
gravity~\cite{Ivanenko:1984vf, DeAndrade:2000sf}.} Solving this constraint equation one finds again the relation \eq{omegaofe0}.



\section{Superfluids}\label{superfluid}

%


A zero-temperature superfluid is a system with a finite density of a spontaneously broken global $U(1)$ charge $Q$. From our perspective, it is an interesting example of the interplay between spontaneously broken internal and space-time symmetries. As such, it is instructive to see how the coset construction is able to reproduce the correct coupling with gravity. Since the low energy description of a superfluid is well known~\cite{Son:2002zn}, and a derivation based on the coset construction was already discussed in~\cite{PhysRevD.89.045002} in the absence of gravity, we will keep this section fairly short and focus mainly on the key technical details.  

The ground state of a superfluid breaks local boosts, time translations and the global $U(1)$ symmetry, but is invariant under the action of $\bar P_0\equiv P_0+\mu Q$, where $\mu$ is the chemical potential~\cite{Nicolis:2011pv}. Thus, from a coset construction perspective, the pattern of symmetry breaking associated with a superfluid is as follows~\cite{PhysRevD.89.045002}:
\bea
\begin{array}{rcl}
\mbox{Unbroken} &=&  \left\{
\begin{array}{ll}
\bar P_0\equiv P_0+\mu Q  &  \qquad \mbox{time translations} \\
\bar P_i \equiv P_i  &  \qquad \mbox{spatial translations} \\
J_{ij} & \qquad \mbox{spatial rotations}
\end{array}
\right. 
\\ && \\
\mbox{Broken} &=&  \left\{
\begin{array}{ll}
K_i \equiv J_{0i} & \qquad\qquad\,\,\,   \mbox{boosts} \\
Q &  \qquad\qquad \,\,\,   \mbox{internal shifts}\, , \\
\end{array}
\right. 
\end{array}
\eea
Therefore, the coset representative can be chosen as
\begin{equation}\label{OmegaSuperfluid}
\Omega = e^{iy^a \bar P_a}  e^{i \pi Q} e^{i\eta^i K_i} \equiv  e^{iy^aP_a}  \, \widetilde{\Omega} \, .
\end{equation}
%
The relevant low-energy degrees of freedom are contained in the covariant version of the Maurer-Cartan form which, using \eq{SuperMaurerCartan}, can be written as
\be\label{superfluidMC}
\begin{split}
\Omega^{-1}D_\mu \Omega 
	&=  \widetilde{\Omega}^{-1}\left(\partial_\mu +i e_\mu^a P_a + \frac{i}{2} \omega_\mu^{ab} J_{ab}\,\right) \widetilde{\Omega} \\
	&= i e_\mu{}^b\Lambda_{b}^{~a} \bar P_a  + i (\partial_\mu \psi -  \mu e_\mu{}^b\Lambda_{b}{}^{0} )Q +	
	\frac{i}{2} J_{ab} \left [  (\Lambda ^ {-1} \partial _ \mu \Lambda) ^ {a b} 
	+ \omega _ \mu ^ {c d} \Lambda _ c \, ^ a \Lambda _ d \, ^ b \right ]\, \\
	&\equiv  i E_\mu{}^a \left ( \bar P_a  +  \nabla _ a \pi \, Q +\nabla _ a \eta ^ i  K _ i +	
	\frac{1}{2} J_{i j} A ^ {ij} _ a \right ) ,
\end{split}
\ee
where in the first equality we have used \eq{SuperMaurerCartan}, while in the second we have introduced the field $\psi \equiv \mu y^0 + \pi$ and the boost matrix
\be
\Lambda^a \, _ b (\eta)  \equiv (e^{i \eta^i K_i} )^a{}_b\,.\label{Lambda_eta}
\ee
\eq{superfluidMC} contains all the building blocks of the low-energy Lagrangian. In particular, one can see immediately that the ``coset vierbein'' is given by $E_\mu{}^a \equiv e_\mu{}^b\Lambda_{b}^{~a}$, and then read off the covariant derivatives for the Goldstones $\pi$ and $\eta^i$:
\begin{equation}\label{superfluidBuildBlocks2}
\nabla_a\pi\equiv e^\mu{}_b\Lambda^{b}_{~a}\partial_\mu \psi - \mu \delta_a^0,\qquad
 \nabla_a \eta^i \equiv e^\mu{}_b\Lambda ^{b}_{~a} \left [  (\Lambda ^ {-1} \partial _ \mu \Lambda) ^ {0 i} 
	+ \omega _ \mu ^ {c d} \Lambda ^ {~0} _ c \Lambda ^ {~i} _ d \right ].
\end{equation}

It is possible to check explicitly that $\nabla_0 \pi$  transforms as a singlet, $\nabla_i \pi$ and $\nabla_0 \eta^i$ as triplets, and  $\nabla_j \eta^i$ as a ${\bf 1}\oplus {\bf 3}\oplus {\bf 5}$  under the unbroken $SO(3)$, and that all are singlets under  diffeomorphisms. Moreover, following the discussion in Section \ref{coset construction}, we conclude that the field
\be\label{superfluidBuildBlocks3}
A _a =  e^ \mu{}_ b \Lambda _ {~a} ^ b \left [  (\Lambda ^ {-1} \partial _ \mu \Lambda) ^ {i j} 
	+ \omega _ \mu ^ {c d} \Lambda ^ {~i} _ c \Lambda ^ {~j} _ d \right ] J_{i j} 
\ee
behaves like  a connection of the $SO(3)$ unbroken group and can be used to define covariant derivatives of matter fields as well as higher covariant derivatives of the Goldstones. 

Equations (\ref{superfluidBuildBlocks2}) and (\ref{superfluidBuildBlocks3}) are the necessary ingredients to write down an effective Lagrangian for superfluids. It is well known however that the low-energy description of superfluids contains a single degree of freedom~\cite{Son:2002zn}. From a coset construction perspective, this result is recovered by noticing that the boost Goldstones $\eta^i$ can be removed from the low-energy spectrum of excitations by imposing the appropriate inverse Higgs constraints~\cite{PhysRevD.89.045002}. In fact, the commutator between unbroken spatial translations and broken boosts gives $[\bar P _i, K_j ] \subset i \delta_{ij} \mu Q$. Based on the discussion in Section~\ref{coset construction}, this means that we can set to zero the spatial covariant derivatives of the Goldstone of $Q$ and solve this constraint to express the $\eta$'s in terms of derivatives of $\pi$:
\be\label{superfluidinversehiggs}
0=\nabla_i \pi = \Lambda  _ {~i} ^ c \left(e_c{}^\mu\partial_\mu \psi\right)\,,\quad \Rightarrow \quad \beta_i = -\frac{e_i{}^\mu\partial_\mu\psi }{e_0{}^\mu\partial_\mu\psi}\,,
\ee
where we have introduced for simplicity the velocity\footnote{With our conventions, the components of the boost matrix $\Lambda^a{}_b$ can be expressed in terms of the velocity $\beta^i$ as follows:
$$
\Lambda^0 \, _0=\gamma\, , \quad \Lambda^0 \, _i=\gamma \beta_i\, , \quad \Lambda^i \, _0= \gamma \beta^i \, , \quad \Lambda^i \, _j=\delta^i\, _j +(\gamma-1)\frac{\beta^i \beta_j}{\beta^2} .
$$}
\be
\la{eta_def}
\beta_i\equiv  \frac{ \eta_i}{\eta} \tanh\eta \, .
\ee
By plugging this result into $\nabla_0 \pi$ we obtain the following lowest order building block of the effective Goldstone boson action:
\be\label{superfluidnabla0}
\nabla_0 \pi =  \Lambda^{~c} _ 0 e_c{}^\mu\partial_\mu \psi  -\mu
= \sqrt{- \eta^{ab}e_a{}^\mu e_b{}^\nu \partial_\mu\psi\partial_\nu\psi} - \mu = \sqrt{ - g^{\mu\nu}\partial_\mu\psi\partial_\nu\psi} - \mu, 
\ee
where in the last step we introduced the inverse space-time metric $g^{\mu\nu}\equiv e^\mu{}_a e^\nu{}_b\eta^{ab}$ \cite{weinberg:1972bo}. Then, the measure of integration $\dd^4 x \det E=\dd^4 x \sqrt{-g}$ is invariant under diffeomorphisms and therefore the relevant Lagrangian at low energies is given by:
\be\label{equ:superfGR1}
S=\int \dd^4 x \det E \  [ a_0 + a_1 \nabla_0 \pi+ a_2 (\nabla_0 \pi)^2 + \cdots]  = \int \dd^4 x \sqrt{-g}  \  F(\sqrt{ - g^{\mu\nu}\partial_\mu\psi\partial_\nu\psi}) \,.
\ee
The function $F$ was introduced in the last step in order to match the more standard notation in flat space-time \cite{Son:2002zn}. This function is defined by the requirement that its $n$-th derivative evaluated at $\mu$ is equal to $a_n$, i.e. $F^{(n)} (\mu) = a_n$. 

As already emphasized in the previous sections, the advantage of the coset construction hinges on the systematics of the derivative expansion. Indeed, higher order terms  can be easily constructed from Eqs.~(\ref{superfluidBuildBlocks2}) and (\ref{superfluidBuildBlocks3}). In particular, the first higher derivative corrections to the low-energy effective action \eq{equ:superfGR1} are $\nabla_0\nabla_0\pi$ and $\nabla_i \eta^i$. After lengthy but straightforward calculations, one can show that these terms can also be written solely in terms of $\psi$ and its derivatives, and in particular
\be
\nabla_0\nabla_0\pi = \frac{\partial_\mu\psi\partial^\mu \d_\rho \psi \d^\rho \psi}{2 \d_\lambda \psi \d^\lambda \psi}, \qquad \qquad \nabla_i\eta^i = - \frac{\left( \d_\rho \psi \d^\rho \psi\Box\psi + \tfrac{1}{2}\partial_\mu  \d_\rho \psi \d^\rho \psi\partial^\mu \psi\right) }{( -\d_\lambda \psi \d^\lambda \psi)^{3/2}}.
\ee
From the perspective of \cite{Son:2002zn} these are just particular linear combinations (with some normalization) of the expected additional higher derivative term.



\section{Membranes}\label{membranes}

We will now use the coset construction to re-derive the effective action for a $d-1$ brane in $(d+1)$ dimensions \cite{Sundrum:1998sj}. The same procedure can be used for higher-codimension branes~\cite{deRham:2007pz,Gomis:2012ki}, extended to superbranes~\cite{Ivanov:1999fwa, Ivanov:1999gy}, and was employed in Ref.~\cite{Brugues:2004an} to describe non-relativistic branes and strings as objects that break the Galilei group. In this section only, our convention for the indices will differ from the one used in the rest of the paper: we adopt a notation that has become  standard in the literature on extra-dimensions (see for instance Ref.~\cite{Sundrum:1998sj}),
\begin{itemize}
\item $A, B, C, D, ...$ will indicate \emph{Lorentz} indices in $d+1$ dimensions.
\item $M, N, P, Q, ...$ will indicate \emph{space-time} indices in $d+1$ dimensions. 
\item $\alpha, \beta, \gamma, \delta, ...$ will indicate \emph{Lorentz} indices in $d$ dimensions.
\item $\mu,\nu,\rho,\sigma, ...$ will indicate \emph{space-time} indices in $d$ dimensions. 
\end{itemize}

For simplicity, let us start  by neglecting gravity---and so for the moment we will not differentiate between Lorentz and space-time indices---and consider the fluctuations about a flat brane. A static brane breaks spatial translations in the direction perpendicular to the brane and Lorentz transformations that mix coordinates on the brane with coordinates in the bulk. Therefore, we can parametrize the coset associated with this symmetry breaking pattern as
\be \la{coset brane}
\Omega = e^{i y^\alpha (x) P_\alpha} e^{i \pi (x) P_d} e^{i \xi^\alpha (x) J_{\alpha d}} \equiv e^{i Y^A (x) P_A} e^{i \xi^\alpha (x) J_{\alpha d}},
\ee
where we find it convenient to introduce the $(d+1)$-dimensional vector $Y^A (x) = ( y^\alpha(x) , \pi(x) )$. These functions describe the familiar embedding of the brane in the bulk, once the reparametrization invariance of the brane is fixed by demanding that the coordinates on the brane are aligned with $d$ coordinates in the bulk. Using the coset parametrization of \eq{coset brane}, we can write the Maurer-Cartan form as:
\begin{subequations} \la{MC brane}
\bea
\Omega^{-1} \d_\mu \Omega &=& i \d_\mu Y^A  \Lambda_A{}^B (\xi) P_B + \fr{i}{2} (\Lambda^{-1})^A{}_C \, \d_\mu \Lambda^{CB} J_{AB} \la{MC brane 1} \\
&\equiv& i E_\mu{}^\alpha \l( P_\alpha + \nabla_\alpha \pi P_d + \nabla_\alpha \xi^\beta J_{\beta d}\r) + i A_\mu{}^{\alpha\beta} J_{\alpha\beta} ,
\eea
\end{subequations}
where $\Lambda_A{}^B (\xi)$ denotes a bulk Lorentz transformation parametrized by the Goldstones $\xi^\alpha$.  The commutation relations $\l[ J_{\alpha d}, P_\beta \r] = i \eta_{\alpha\beta} P_d $ tell us that at low energies to impose the inverse Higgs constraint $\nabla_\alpha \pi \equiv 0$ to express the Goldstones $\xi^\beta$ in terms of derivatives of~$\pi$. As in the previous example, the covariant derivative $\nabla_\alpha \xi^\beta$ will enter the action only at higher order in the derivative expansion. Thus, at lowest order in derivatives, the effective action for a brane is 
\bea
S &=& -T \int d^d x \det E = -T \int d^d x \sqrt{- \det (E E^T) \det (\eta)} = \nonumber  \\
&=& -T \int d^d x \sqrt{- \det ( \d_\mu Y^A  \Lambda_A{}^\gamma \d_\nu Y^B  \Lambda_B{}_\gamma )} = -T \int d^d x \sqrt{ - \det ( \d_\mu Y^A  \d_\nu Y_A )}  \qquad \la{Eff Act brane no gravity}
\eea
where $T$, the brane tension, is an energy per unit area and we have used the fact that the inverse Higgs constraint  $\nabla_\alpha \pi = 0$ implies $ \d_\mu Y^A  \Lambda_A{}^d=0$.

While our approach has been purely algebraic, it maps nicely to the usual geometric interpretation. To begin with, \eq{Eff Act brane no gravity} reproduces a familiar result: the low-energy effective action for a brane is given by the square root of the determinant of the induced metric
\be \la{brane ind metr}
h_{\mu\nu} \equiv \eta_{AB} \d_\mu Y^A \d_\nu Y^B = \eta_{\mu\nu} + \d_\mu \pi \d_\nu \pi.
\ee
Furthermore, from the geometrical point of view, the constraint $\d_\mu Y^A  \Lambda_A{}^d=0$ identifies $\Lambda_A{}^d(\xi)\equiv n_A$ as a unit vector perpendicular to all the $\d_\mu Y^A\,$s and therefore to the surface itself. Supplied with this unit vector we can calculate its change as we move around the world volume projected on the vectors tangent to the world volume---this is the extrinsic curvature. After some manipulations one can show that the higher derivative covariant objects, $\nabla_\alpha \xi_\beta$, are proportional to the extrinsic curvature:
\be
\nabla_\alpha \xi_\beta=E^\mu{}_\alpha E^\nu{}_\beta \d_\mu Y^A \d_\nu Y^B \d_A n_B= E^\mu{}_\alpha E^\nu{}_\beta K_{\mu \nu}.
\ee

Similarly, to compute derivatives along the world volume one has to take into account the spin connection associated with the induced metric. This should be precisely related to the covariant derivative (\ref{covdev}) supplied by the algebraic construction, and indeed it is easy to show that this is the case. In this sense, there is a one-to-one mapping between the algebraic objects constructed above and the more standard geometrical ones of the extrinsic curvature, the induced metric and its spin connection. For the interested reader a more lengthy and explicit discussion can be found in Appendix \ref{geometry for membrane}.

\subsection{Coupling with Gravity}

We can now introduce gravity in the bulk, in the language of the coset construction, following Section \ref{gravity_and_coset}. First, now that we are dealing with curved space, we must differentiate the position of the membrane in the local Lorentz frame and the global space-time. We do so with $Y^A(x)$ and $Y^M(x)$ respectively. Proceeding in several steps, we first parametrize the coset as
\be
\la{coset brane grav}
\Omega =  e^{i Y^A (x) P_A} e^{i \xi^\alpha (x) J_{\alpha d}},
\ee
and then rewrite the covariant version of the Maurer-Cartan form as follows:
\be
\Omega^{-1} D_\mu \Omega \equiv \d_\mu Y^M \Omega^{-1} D_M \Omega,
\ee
where we have expressed the derivatives along the coordinates in the brane as projected derivatives of the space-time coordinates, which include the Poncar\'e gauge fields, as discussed in Section \ref{gravity_and_coset}. Then,
\bea
\d_\mu Y^M \Omega^{-1} D_M \Omega &=&  i \d_\mu Y^M e_M{}^A \Lambda_A{}^B P_B + \fr{i}{2} (\Lambda^{-1})^A{}_C ( \eta^{CD} \d_\mu + \d_\mu Y^M \omega_M^{CD})\Lambda_D{}^B J_{AB} \, \nn \\
&=&  i E_\mu{}^\alpha \l( P_\alpha + \nabla_\alpha \pi P_d + \nabla_\alpha \xi^\beta J_{\beta d}\r) + i A_\mu{}^{\alpha\beta} J_{\alpha\beta}  \label{brane + gravity}
\eea
As in Section \ref{gravity_and_coset}, we defined $e_M{}^A \equiv \d_M Y^A + \tilde{e}_M{}^A + \tilde{\omega}_M^{AC} Y_C $ and $ \tilde{\omega}_M^{AC} = \omega_M^{AC} $. Here however, they are evaluated on the membrane itself.
By comparing Eqs.~(\ref{brane + gravity}) and (\ref{MC brane 1}), we see that the coupling with gravity modifies the results we obtained previously in two ways:
\begin{enumerate}
\item it replaces $\d_\mu Y^A$ with $\d_\mu Y^M e_M{}^A$
\item it replaces every partial derivative $\d_\mu$ in $\nabla_\alpha \xi_\beta$ and $A_\mu{}^{\alpha\beta}$ with $\d_\mu + \d_\mu Y^M \omega_M$,
\end{enumerate}
which  is what one could have guessed by examining \eq{Eff Act brane no gravity}. The low-energy effective action now becomes  
\be
S =  -T \int d^d x \sqrt{ - \det ( \d_\mu Y^M e_M{}^A  \d_\nu Y^N e_{NA} )} = -T \int d^d x \sqrt{ - \det ( h_{\mu\nu} )} \, .
\ee
Similar to our discussion in the flat space case, the higher order covariant derivatives can be related to the extrinsic curvature
\bea
\nabla_\alpha \xi_\beta &=& E^\mu{}_\alpha E^\nu{}_\beta\d_\mu Y^M e_M{}^A \d_\nu Y^N e_N{}^B \nabla_A n_B =   (\Lambda^{-1})_\alpha{}^C (\Lambda^{-1})_\beta{}^B  \nabla_C n_B \nonumber \\
 &=& E^\mu{}_\alpha E^\nu{}_\beta K_{\mu \nu} \; ,
\eea
where $n_A \equiv \Lambda_A{}^d(\xi)$ is again, by the constraint equation, the unit normal vector perpendicular to the membrane in the local Lorentz frame.

Furthermore, the covariant derivative of matter fields living on the brane is now
\be
\nabla_\alpha \psi = (E^{-1})_\alpha{}^\mu \l[\d_\mu \psi + \fr{i}{2} (\Lambda^{-1})^\beta{}_B \, (\eta^{BC} \d_\mu +  \d_\mu Y^M \omega_M^{BC})\Lambda_C{}^\gamma J_{\beta\gamma} \psi \r] \, .
\ee
One can show that the connection term  $(\Lambda^{-1})^\beta{}_B \, (\eta^{BC} \d_\mu +  \d_\mu Y^M \omega_M^{BC})\Lambda_C{}^\gamma$ is indeed the spin connection associated with the induced metric. Hence, we see that the one-to-one correspondence between the objects generated by our algebraic approach and the usual geometrical one persists even when the bulk geometry is curved.



\section{Point Particles} \la{point particles}

{\gre }

In this section we describe a free pointlike particle coupled to gravity, using coset-construction techniques. This is of course the  limiting case of the low energy theory for a general membrane developed in the preceding section, but we report it here for two reasons. First, for its simplicity and its easy interpretation in terms of familiar physics. Second, because it will serve  as an opportunity to develop the notation for the case of spinning point-like objects which we discuss in the next section. 

The symmetry breaking pattern for the point particle can be read from the membrane case of section \ref{membranes}, in the limit where the brane is one-dimensional and oriented in the time-like direction: 
\be
\label{pattern for point part}
\begin{array}{rcl}
\mbox{Unbroken} &=&  \left\{
\begin{array}{ll}
P_0  &  \,\qquad\quad\quad \mbox{ time translations} \\
J_{ij}  & \,\qquad\quad\quad \mbox{ spatial rotations}
\end{array}
\right. 
\\ && \\
\mbox{Broken} &=&  \left\{
\begin{array}{ll}
P_i &  \quad\quad   \mbox{spatial translations} \\
J_{0i}\equiv K_i & \quad\quad   \mbox{boosts},
\end{array}
\right. 
\end{array}
\ee
and all translations are non-linearly realized, as discussed in Section~\ref{gravity_and_coset}. We parametrize our coset by
\be
\label{coset parametrization for point part}
\Omega= e^{i y^a (\lambda) P_a}e^{i\eta^i(\lambda)K_i} \equiv  e^{i y^a(\lambda) P_a} \tilde\Omega \;,
\ee
where $\lambda$ is some monotonic parameter that traces out the worldline of the particle. Just as in the membrane case, the covariant version of the Maurer-Cartan form projected onto the particle's worldline is:
\bea
\label{M-C form}
\dot x^\mu\Omega^{-1}D_\mu \Omega&=&{\dot x^\mu}\tilde{\Omega}^{-1}\left(\di_\mu +i e_\mu{}^a P_a +\frac{i}{2}\omega_\mu{}^{ab}J_{ab} \right)\tilde{\Omega} \\
&\equiv&i E(P_0+\nabla \pi^i P_i +\nabla \eta^i K_i+A^{ij}J_{ij})\;. \nonumber
\eea
and the dot denotes a derivative with respect to $\lambda$. Explicit computation gives
\bea
E&=&\dot  x^\nu e_\nu{}^a \Lambda_{a}\, ^0\label{nablaetapp0} \\
\nabla \pi^i&=& E^{-1} \dot x^\nu e_\nu{}^a \Lambda_{a}\, ^i\\
\nabla \eta^i&=& E^{-1} \left((\Lambda^{-1})^0{} _c \dot \Lambda^{c i} +\dot x^\mu\omega_\mu {}^{ab} \Lambda_a{}^0 \Lambda_b{}^i\right)\label{nablaetapp} \\
A^{ij}&=& \f{E^{-1}}{2} \left((\Lambda^{-1})^i{} _c \dot \Lambda^{c j}+\dot x^\mu\omega_\mu {}^{ab} \Lambda_a{}^i \Lambda_b{}^j  \right)\label{nablaetapp1}
\eea
where the boost matrix $\Lambda^a{}_b\equiv \Lambda^a{}_b(\eta)$ is a function of the Goldstone bosons defined in Eq.~(\ref{Lambda_eta}). As discussed in Section~\ref{coset construction}, we can deduce the existence of an inverse Higgs constraint from the fact that the commutator between unbroken time translations and boosts gives broken spatial translations. We can therefore set to zero the covariant derivative
\be 
\label{constraint eqn for free point part}
\nabla \pi^i=E^{-1}\left( \dot x^\nu e_\nu{}^0\Lambda_0 \,^i+ \dot x^\nu e_\nu{}^j\Lambda_j \,^i\right)=0 \; .
\ee
Expressing the boost matrix in terms of velocities, as defined in \eq{eta_def}, this equation takes the simple form:
\be
\label{solution to free part const}
\beta^i=\frac{\dot x^\nu e_\nu{}^i}{\dot x^\nu e_\nu{}^0}\,.
\ee
In flat space-time, one can choose coordinates such that $e_\nu{}^a=\delta_\nu{}^a$, and the constraint above gives $\beta^i=\partial_0 x^i$. The physical interpretation of this solution is then clear:  $\vec \beta$ (or equivalently $\vec \eta$) parametrizes the boost necessary to get into the moving particle rest frame. A similar interpretation holds in curved space, as will become clear in what follows.

From  equations~(\ref{nablaetapp0})--(\ref{nablaetapp}) we see that,
\bea
| E | &=& \sqrt{E^2}=\sqrt{(E\nabla \pi^i)^2-(\dot x^\nu e_\nu{}^a \Lambda_{a}\, ^c \dot x^\mu e_\mu{}^b \Lambda_{b c})}= \sqrt{-(\eta_{ab} e_\nu{}^a e_\mu{}^b\dot x^\nu \dot x^\mu  )}\nn\\
&=&
\sqrt{-g_{\mu \nu} \dot x^\mu\dot x_\mu} \equiv\frac{d\tau}{d\lambda},
\eea
where in the third equality we have utilized the constraint (\ref{constraint eqn for free point part}). This result allows us to rewrite the constraint itself in a way that makes its physical interpretation  manifest:
\be\label{constraintwithU}
u^a\Lambda_a{}^i(\eta) =0,
\ee
where $u^a\equiv e_\mu{}^a \partial_\tau x^\mu $ is the Lorentz velocity as measured in the local inertial frame defined by the vielbein (in the flat space case, with $e_\mu{}^a=\delta_\mu^a$, this reduces to the usual definition of the four-velocity). 

Similarly to the membrane case, as discussed in the previous section as well as in Appendix \ref{geometry for membrane}, there is a simple geometrical interpretation of the quantities defined above. 
The set of  local Lorentz vectors 
\be
\label{geometrical basis in Lorentz indices pp}
\left\{\hat{n}^{a}{}_{(0)}\equiv u^a= \Lambda^a{}_0(\eta)\, , \quad\, \hat{n}^{a}{}_{(i)}\equiv \Lambda^{a}{}_{i}(\eta)\right\},
\ee
define an orthonormal (w.r.t. the local Minkowski metric defined by the vielbeins) basis in the comoving frame of the particle, where orthogonality follows from the property of boost matrices, $\Lambda_a{}^b\Lambda^a{}_c=\delta^b_c$. In other words, the $\hat{n}^{a}{}_{(b)}$ is itself a set of vielbeins that defines the local inertial frame on the particle trajectory (alternatively, the set $\hat{n}^{\mu}{}_{(b)}\equiv e^\mu{}_a\hat{n}^{a}{}_{(b)}$ is orthonormal w.r.t. the full metric $g_{\mu\nu}$ and defines an orthonormal comoving basis in terms of space-time vectors -- with index $\mu$). Now, the covariant derivatives  $\nabla^i\eta$ of the boost Goldstones given in  \eq{nablaetapp} can be rewritten as
\be\label{etapp}
\nabla \eta^i =\hat n_{a}{}^{ (i)} \left(\di_\tau u^a+u^\mu\omega_\mu{}^a{}_c u^c \right)=
\hat n_{a}{}^{ (i)} u^\mu \nabla _\mu u ^ a = \hat n_{a}{}^{ (i)}e_\mu{}^a a^\mu,
\ee
where the $\nabla _\mu$ introduced in the second step is the usual covariant derivative of GR, and in the third step we have used the standard definition of the acceleration $ a^\mu \equiv \partial _ \tau u ^ \mu + \Gamma ^ \m _ {\lambda \sigma} u ^ \lambda u ^ \sigma$. As one can see, the physical meaning of the covariant derivatives  $\nabla \eta^i$ is that they correspond to the component of the acceleration $a^a\equiv e_\mu{}^a a^\mu$ (as measured in a local inertial frame on the particle trajectory) projected on the $i$-th vector of the basis defined in \eq{geometrical basis in Lorentz indices pp}. From a geometrical point of view, the $\nabla\eta^i$ correspond to the extrinsic curvature of the world line, defined as the covariant derivative of the normal vectors projected onto the worldline:
\be \la{ext curv point w grav}
K^{(i)} \equiv \d_\tau x^\nu (\partial _\tau \hat n_\nu^{(i)} + \Gamma _ {\nu, \lambda \sigma} \d_\tau x ^ \lambda \hat{n}^{\sigma \, (i)}) =  
u ^ \nu e ^ a _ \nu \nabla _ \tau \hat n _ a ^ {~(i)} = u ^ a  \nabla _ \tau \hat n _ a ^ {~(i)} = - \nabla \eta ^ i \, ,
\ee
where in the last step we have utilized the constraint (\ref{constraintwithU}). 

We are now ready to build the leading order action for the point particle using the covariant objects of Eqs.~(\ref{nablaetapp0}) -- (\ref{nablaetapp1}). Since $\nabla \pi^i=0$, $\nabla\eta^i$ is a higher derivative term and there is nothing in the field content to build any leading order objects with the covariant derivative formed from $A_{ij}$, we are left with the simple action
\be\label{Eff Act pp with gravityxx}
S = - m \int d \lambda \, E= -m \int d\tau  \;,
\ee
which matches  the well known expression for the action for the point particle, when the dimensionful coefficient $m$ is identified with the particle's mass.\footnote{ From the perspective of the previous section, the low-energy effective action is given by the square root of the determinant of the induced (in this case $1$-dimensional) metric on the particle trajectory: $ h_{00} = g_{\mu \nu} \dot x^\mu \dot x^\nu$. 
Then, the ``coset einbein'' $e_0{}^0\equiv E$ behaves really like a einbein for the induced metric, i.e.
$ h_{00} = e_0{}^0  e_0{}^0 \eta_{00}$.}
It is interesting to note that, in the absence of external fields or gravity, the lowest order equations of motion are equal to $a^\mu=0$, and so higher-derivative terms proportional to $\nabla\eta^i$ in the full action are, in fact, proportional to the lowest order equations of motion. This means that they will not contribute to any physical observable and can simply be removed by a field redefinition, implying that \eq{Eff Act pp with gravityxx} is the correct action at \emph{all} orders to describe the free point particle. This seems odd from an EFT perspective as one would expect a tower of terms in the low energy Lagrangian that encode effects of the integrated out UV-physics.\footnote{The mass of the object is an IR quantity that can be measured for a given particle at infinity (in asymptotically flat space) and carries no information of the UV-physics: therefore it cannot be related with the scale appearing in the EFT expansion. Indeed, the (ADM) mass of the object is defined at infinity and so there is no way to tell the difference between  a black hole of one earth mass or the earth itself.} This peculiarity is an accident of the simplicity of our construction and it follows from the fact that the extrinsic curvature is related to the equations of motion by equations (\ref{etapp}) and (\ref{ext curv point w grav}). This is however not the case for higher dimensional objects such as membranes, for which the extrinsic curvature gives rise to physical effects. Higher derivative terms in the action can also appear in the effective action by adding external fields (for instance with the inclusion of non-minimal couplings to gravity~\cite{Goldberger:2004jt}) or when orientational (spin) degrees of freedom are taken into account, as we will discuss in the next section.

When we do include gravity, there are indeed operators that can be added to Eq.~(\ref{Eff Act pp with gravityxx}) that encode the finite size extent of the point particle and are absolutely necessary from an EFT point of view \cite{Goldberger:2004jt}. As discussed in Section \ref{gravity_and_coset}, with gravity in the picture we have the additional field given by the Riemann curvature tensor $R^{ab}{}_{cd}$. However, at the moment $R^{ab}{}_{cd}$ transforms linearly under Lorentz, a symmetry which the point particle instead realizes nonlinearly. We can remedy this situation by defining, schematically,
\be
\tilde{R} \equiv \Omega^{-1}_L(\pi) \cdot R \;.
\ee
Where $\Omega_L(\pi)=\tilde{\Omega}$ is the Lorentz part of Eq.~(\ref{coset parametrization for point part}). One can easily check that $\tilde{R}$ transforms under a Lorentz transformation as $\tilde{R} \rightarrow h \tilde{R}$ where $h$ is an element of the unbroken rotation group. Explicitly writing out the indices we have
\be
\label{R tilde}
\tilde{R}^{ab}{}_{cd}=\left(\Lambda^{-1}(\eta)\right)^a{}_e \left(\Lambda^{-1}(\eta)\right)^b{}_f \left(\Lambda^{-1}(\eta)\right)_c{}^g \left(\Lambda^{-1}(\eta)\right)_d{}^h R^{ef}{}_{gh}
\ee
where the $\eta$'s in the boost matrices are, of course, those satisfying the inverse Higgs constraint given by Eq.~(\ref{constraint eqn for free point part}). Or more physically, $\tilde{R}$ is simply the Riemann curvature tensor in the local rest frame of the moving particle. As the reader familiar with the coset construction may have already noticed, this is just the usual procedure used to dress ``matter fields'' that transform in a linear representation of the full group $G$  into fields that transform in a linear representation of the unbroken group $H$ \cite{Weinberg:1996kr}.

Furnished with these correctly transforming fields we can now form rotationally invariant objects out of $\tilde{R}$ and integrate them along with our invariant measure. In particular, we have that
\be
\tilde{R}_{00}=u^\mu u^\nu R_{\mu \nu}  \quad \text{ while } \quad \tilde{R}_{ii}= R+u^\mu u^\nu R_{\mu \nu}\; .
\ee
These terms describe finite size effects, as they are proportional to the curvature variation on scales given by the size of the object \cite{Goldberger:2004jt}.\footnote{Where we instead to consider a point particle coupled to a $U(1)$ gauge field we could have begun instead with the field strength tensor $F_{ab}$. Upon application of $\tilde{\Omega}^{-1}$ we would have generated an appropriately transforming $\tilde{F}$. Finite size terms would take the form of rotationally invariant contractions such as $\tilde{F}_{0i} \cdot \tilde{F}_{0i}$ which can be written in the explicitly Lorentz invariant fashion as $u^\mu u^\nu F_{\mu \sigma}F_{\nu}{}^\sigma$. }


\section {Spinning Objects} \la{spinning particle}
What is the symmetry breaking pattern of a generic finite-size object, like a black hole or a lumpy asteroid? Generically one expects that the full Poincar\'e group is now broken and the system can be described similarly to the point particle case but with the additional breaking of rotations.\footnote{A similar point of view was adopted in Ref.~\cite{Papenbrock:2013cra} to derive an effective theory for atomic nuclei.} However, extended objects can have their own additional symmetries, and these must be taken into account to properly characterize the behavior of the system at low-energy. We will refer to this additional symmetry group $S\subseteq SO(d)$ as an \emph{internal} symmetry so that $G=ISO(3,1)\times S$, is the fundamental global symmetry of the system. For example, $S=SO(d)$ for a $(d-1)$-dimensional sphere, while $S=\emptyset$ for a lumpy asteroid. Notice that $S$ could also be a discrete group, and this would be appropriate to describe a regular polyhedron.  Our choice of coset parametrization will be such that we can seamlessly treat both continuous and discrete internal symmetries.

In the rest frame of the object, $G$ is broken down to a linear combination of internal rotations (with generators $S_{ij}$) and spatial rotations (with generators $J_{ij}$). The symmetry breaking pattern is then the following	
\be
\label{pattern for spinning general}
\begin{array}{rcl}
\mbox{Unbroken} &=&  \left\{
\begin{array}{l}
P_0 \\
\bar{J_{ij}}   
\end{array}
\right. 
\\ && \\
\mbox{Broken} &=&  \left\{
\begin{array}{l}
P_i \\
J_{ab} 
\end{array}
\right. 
\end{array}
\ee
where $\bar{J}$ is the unbroken linear combination of the internal and space-time rotations. For instance, for a spherical object $\bar{J}_{ij}=S_{ij}+J_{ij}$ where $S_{ij}$ are the generators of the internal $SO(d)$ group. The coset can be parametrized by
\be
\label{coset parametrization for general spin}
\Omega=e^{i y^a P_a} e^{{i}\alpha_{ab} J^{ab}/2 }=e^{i y^a P_a }e^{i\eta^i J_{0i}}e^{i \xi_{ij}J_{ij}/2}\,,
\ee
where in the second equality we have used the fact that any Lorentz transformation can be written \emph{uniquely} as the product of a rotation and a boost, 
implying a one-to-one correspondence between the Goldstone fields  $\alpha_{ab}$ and their alternative representation as $\eta_i,\xi_{ij}$. Notice that we chose to define our coset parametrization (\ref{coset parametrization for general spin}) using the generators of broken spatial rotations (as opposed to the internal ones). This choice is particularly convenient because it allows us to calculate the Maurer-Cartan form without specifying the exact form of the unbroken generators $\bar{J}_{ij}$.

The relevant degrees of freedom can now be identified by projecting the covariant Maurer-Cartan form on the worldline of the object.  Similarly to \eq{M-C form}, this can be written as
\be
\label{M-C form for general spin}
\dot x^\mu \Omega^{-1} D_\mu \Omega=
iE(P_0+\nabla\pi^iP_i +\frac{1}{2}\nabla \alpha_{cd} J^{cd}) \,.
\ee
We can then write the relevant objects that describe the low-energy dynamics explicitly: 
\bea
E&=&\dot  x^\nu e_\nu{}^a \Lambda_{a} \, ^0 \nonumber 
\\ 
\nabla \pi^i&=& E^{-1} \dot x^\nu e_\nu{}^a \Lambda_{a} \, ^ i \label{spincovariants2}\\
\nabla \alpha^{ab}&=&   E^{-1} \left( \Lambda _ c ^{~a} \dot\Lambda^{c b} + \dot{x}^ \mu \omega_\mu {}^{c d} \Lambda_c{}^a \Lambda_d{}^b \right), \nonumber 
\eea
where here $\Lambda$ is once again a Lorentz transformation either parametrized by $\alpha$ or, equivalently, by $\eta$ and $\xi$. On general grounds, one would have also expected a connection term proportional to $\bar{J}$ on the right hand side of \eq{M-C form for general spin}. As alluded to earlier, one of the benefits of the coset parametrization (\ref{coset parametrization for general spin}) is precisely that such a connection will not appear. Moreover, it is worth stressing the fact that the covariant building blocks (\ref{spincovariants2}) are independent of the residual symmetry group.

The presence of the unbroken rotations will manifest itself in the way we  contract the indices of the objects in \eq{spincovariants2} to build the invariant terms that appear in the Lagrangian.  In general as  $[\bar{J}_{ij},{J}_{kl}]\neq0$  (recall that if $\left\{\bar{J}_{ij} \right\}$ is not empty it will include a non-vanishing contribution from $J_{ij}$) the objects of \eq{spincovariants2} will transform linearly under $H=\{\bar{J}_{ij}\}$. If our residual symmetry is $SO(d)$ then we will contract all the spacial indices in an $SO(d)$-invariant manner. 
For less symmetrical objects there are simply more possibilities restricted only by the form of $H$. This procedure makes it clear how to describe objects with inherent multipole moments (e.g. Saturn's non-spherical moon, Hyperion); we simply contract the indices in the necessary fashion.
 
Just as in the the point-particle case, the low-energy effective theory can be derived by imposing the covariant constraint $\nabla \pi^i=0$. Utilizing the fact that, for rotations, $\Lambda^0 \,_a(\xi)=\delta^0_a$ and $\Lambda^i\, _j (\xi)=R^i\, _j(\xi)$, with $R(\xi)$ an $SO(d)$ matrix, the constraint reads
\be
u^a\Lambda_a{}^i(\eta)R_ j \, ^ i(\xi) =0 \, .
\ee
Since $R^{k}\,_{i}(\xi)$ is invertible, this gives
\be\label{inverse higgs const for spinning in flat space}
u^a\Lambda_a{}^i(\eta) =0.
\ee
This is the same constraint equation that we encountered in the case of the featureless point particle, \eq{constraintwithU}. As such, it can be solved identically and it admits the same geometrical interpretation: the $\Lambda(\eta) ^a \, _ b$ as a set of \emph{local} orthonormal vectors~$\{\hat n^a{}_{(b)}\}$. 
In the spinning case, however, we now have an additional set of orthonormal vectors 
\be \la{Lambda_decomposition}
\hat m ^b{} _{(a)}\equiv\Lambda^b{}_{a}(\alpha)=\Lambda^b \, _c (\eta)\Lambda^c\, _a (\xi).
\ee
The 0-th vector  $\hat m ^b{} _{(0)}=\hat n ^b{} _{(0)}=u^b$ coincides  with the velocity of the particle in the free-falling rest frame, while the other vectors differ by a rotation $\Lambda(\xi)$. That is, the set of vectors $\{\hat m^b{}_{(a)}\}$ encode the additional information of {\em rotation}, paramatrized by the $d(d-1)/2$ independent degrees of freedom of $\xi$.

After imposing the constraint $\nabla \pi^i=0$, it is easy to show that the covariant derivatives $\nabla \alpha^{0i}$ can always be removed from the action order by order in the derivative expansion. Indeed, from \eq{spincovariants2} and (\ref{Lambda_decomposition}) we obtain
\be
\nabla \alpha^{0i}=\Lambda_j{}^i(\xi) \hat n_a{}^{(j)} \nabla _\tau u^a \;,
\ee
which is merely a rotated version of the $\nabla \eta^i$ encountered in the featureless point particle case---see \eq{etapp}. By the same arguments (see below \eq{Eff Act pp with gravityxx}), these terms are proportional to the leading order equations of motion $a^\mu =0$, and as such can be eliminated through a field redefinition. As a consequence, the effective action at lowest order in the derivative expansion reads
\be\label{lumpyaction}
S=\int d\lambda\, E\, \left(-m+ \frac{I_{ijkl}}{4}\nabla \alpha^{ij}\nabla \alpha^{kl}+\cdots\right),
\ee
where  $E$ and $\nabla \alpha$ from \eq{spincovariants2} are evaluated on the solutions of the constraint \eq{inverse higgs const for spinning in flat space}, we have discarded the linear term in $\nabla \alpha$ by time reversal symmetry and the dots denote higher order terms.The explicit form of the coefficients $I_{ijkl}$ are invariant under the unbroken group $H$ and encode the residual symmetries of the object. \footnote{Just as in the point particle case, there will be ``finite size'' terms which can be constructed by first applying $\tilde\Omega^{-1}$ to the Riemann curvature tensor as in (\ref{R tilde}) and then contracting it with itself and the new structure $\nabla \alpha^{ij}$. Here $\tilde R$ differs from the one in the point particle case by an additional rotation contained in the $\Lambda$'s. In this way we can generate the richer set---in comparison to the point particle case---of finite sized terms reported in \cite{Porto}.}

Physically, we expect the coefficients $I_{ijkl}$ to be related to the moments of inertia. This is made most clear by considering the $3+1$ dimensional case where we can define the rotations as vectors via the epsilon tensor, $\theta_i=\frac{1}{2}\epsilon_{ijk} \xi^{jk}$. These can be thought of as the angles describing the instantaneous orientation of our spinning object.\footnote{Notice that these angles are not the usual Euler angles. Our rotation matrix is parametrized as $R(\vec{\theta})\equiv \exp{(i\theta^iJ_i)}$, and thus $\theta^i$ is more precisely the three vector about which a rotation by an angle $| \vec \theta|$ is performed. On the contrary, the standard definitions of the Euler angles decompose the total orthogonal matrix into a product of three rotations around two different axes, such as for instance $R(\alpha,\beta,\gamma)\equiv \exp{(i\alpha J_z)} \exp{(i\beta J_x)} \exp{(i\gamma J_z)}$. } Expressing the above in these variables, the moment of inertia takes the more familiar form with two indices: $\frac{1}{2}I_{ij}\nabla \theta^i \nabla \theta^j$ where $I_{ij}$ is the usual two-index moment of inertia tensor for rotations of arbitrary rigid body in $3+1$ dimensions. The curious reader might then wonder what is the physical interpretation of the coefficients that appear in front of the higher order terms denoted by the dots in Eq.~(\ref{lumpyaction}).

To simplify the discussion, let us consider a spherical object. In this case $H=SO(d)$ and we must contract the indices accordingly, which means that without loss of generality, we can set $I_{ijkl}  \sim I \delta_{ik} \delta_{jl}$. The action is simply
\be
\label{flat space sphere action}
S=\int d\tau \,\left(-m +\frac{I}{4} \, \nabla \alpha^{ij}\nabla \alpha_{ij}+ \cdots \right) \;,
\ee
where the dots stand for higher derivative terms suppressed by some UV scale. How does this action make contact with our usual understanding of rotational dynamics?

From classical mechanics, we know that only two  physical parameters are necessary to describe a completely rigid spherical object: the mass $m$ and the moment of inertia $I$. What degrees of freedom have been neglected by going to the ``completely rigid'' limit? From the effective field theory perspective, as we go deeper into the UV we expect to encounter the degrees of freedom associated with the elasticity of the object: the normal modes. And so, as the action given in Eq.~(\ref{flat space sphere action}) is the action obtained after integrating out all these degrees of freedom. Hence, we expect the higher order terms to be suppressed by inverse powers of the characteristic frequency $\omega_N$ of these modes.

As an explicit example, let's consider a solid material body of large enough size that the surface tension can be neglected but is nevertheless bound by its own intermolecular forces (as opposed to gravitational ones). For instance, a sphere of iron of many meters in diameter.\footnote{ From the effective field theory point of view, this is just a particular ``UV-completion'' of our theory.} This allows us, for simplicity, to neglect gravity and focus only on the rotational degrees of freedom. The typical frequency of this system's normal modes is related to the speed of sound in the material $c_s$ and the typical length scale $L$ of the object by\footnote{The same analysis applies to black holes. There, the characteristic time scale is given by the light crossing time. And so, as would be expected dimensionally, the frequency of the (quasi-)normal modes of a black hole $ \sim c/L$, where $c$ is the speed of light~\cite{Kokkotas:1999bd}, and the $L$ is given by the Schwarzschild radius $2Gm$.}  
\be
\omega_N \sim c_s/L.
\ee
For simplicity, let us restrict ourselves to the case in which the translational velocity is zero and focus on the rotational dynamics. The covariant derivatives $\nabla \alpha^{ij}$ of the angular variables will be of order the rotational frequency $\omega_R$. The leading order piece of the Lagrangian then is of order $\sim I \cdot \omega_R^2 $.

In the simple case we are considering, enforcing time reversal symmetry, there are three possible next to leading order terms given schematically by
\be
\left[\nabla \alpha^4 \right] \;,\quad \left[\nabla \alpha ^2 \right]^2 \quad \text{ and } \quad \left[\left(\tfrac{d}{dt} \nabla \alpha \right)^2 \right]\; ,
\ee
where the brackets denote the trace. Using the lowest order equations of motion, $\tfrac{d^2}{dt^2} R^{ij}=0$, the third term can be rewritten as the second; leaving us with just two possible next to leading order structures. Dimensionally, each of these terms comes with two additional time derivatives in comparison with the leading term. As such, the coefficients accompanying these terms,  let's call them $\Xi$, are down by two powers of the characteristic frequency of the integrated out modes. That is,
\be
\label{higher_order_scaling}
\Xi \, (\nabla \alpha)^4 \sim \fr{I}{{\omega_N}^2} \, \omega_R^4  \sim   \fr{\omega_R^2 L^2}{{c_s}^2} \, I \cdot \omega^2_R\sim \delta I (\omega_R) \, \omega_R^2 .
\ee

The physical interpretation of these higher order terms is clear: they are related to how the body deforms under a finite rate of rotation. These deformations lead to corrections both in the rotational energy and in the energy related to the deformation itself. One can see from the scaling in Eq.~(\ref{higher_order_scaling}) that the tower of higher order terms is under perturbative control as long as the rotational velocity is much less than the speed of sound of the material. For a standard material body, this rotational frequency would be precisely that at which the body would undergo large non-linear stresses and order-one distortions, dramatically exiting the regime of validity of the effective theory we have constructed.



\section{Discussion and Conclusions}

By formulating GR as a gauge theory associated with local Poincar\'e symmetry and the non-linear realization of translations we have been able to seamlessly extend the coset construction to describe the coupling between gravity and systems whose ground state breaks space-time symmetries. We have illustrated the power of our method by constructing the low energy effective actions describing the coupling of gravity to three simple, but important, systems: superfluids, membranes embedded in higher dimensional space, and spinning objects.

The value of the first two examples is mostly pedagogical. The minimal coupling of these systems to gravity is manifest from the formalism introduced in \cite{Son:2002zn, Sundrum:1998sj} and therefore we have shown explicitly that our construction matches exactly the known results. The superfluid case illustrates how gravity couples to systems which break some internal and space-time symmetries down to a diagonal subgroup. In the membrane section we instead showed how the usual geometric picture arises naturally from an algebraic approach such as the coset construction.  We then applied the lessons learned from these examples to the description of spinning objects---perhaps the most interesting application of our techniques. 

Physically, the point-like spinning objects that we consider provide a low-energy description for rotating astrophysical bodies like, for example, planets, black holes and neutron stars. Despite their physical importance, these systems have proven difficult to describe from the low-energy perspective. The crux of the construction is the following. In order to couple a spinning object to gravity one usually starts from a Lorentz covariant theory and, by replacing $\di_\mu \rightarrow \nabla_\mu$ and $\eta_{\mu \nu} \rightarrow g_{\mu \nu}$ introduces a minimal coupling with gravity. The problem of this approach is that the rotational degrees of freedom cannot be captured in a Lorentz covariant way without introducing redundances. In order to recover an action that has the right number of degrees of freedom,  additional constraints must be implemented \cite{Hanson:1974qy}. This  Lorentz covariant approach can then be extended  to describe the curved space case~\cite{Porto}. 

In this article we decided to follow instead a different approach. By considering the spacetime symmetries that are spontaneously broken by a rigid body at rest, we managed to build an effective action which is a natural generalization of that of a non-relativistic rigid body. Such an effective action describes bodies that rotate slowly in their center of mass frame but move at arbitrary speeds. Put another way, the action for the spinning object is resummed already to all orders in the translational velocity, $v/c$, but is organized as a polynomial in powers of the rotational velocity over the speed of sound, $v_R/ c_s$.\footnote{For relativistic matter, $c_s \sim c$.} Neglecting gravity, this makes the non-relativistic limit very clear: when $v\rightarrow 0$ we recover an action for the rotational modes of {\em precisely} the same form as what we would have obtained starting from the Galilei group. 

If one is interested in working in the post-Newtonian approximation where $v/c \ll 1$ one can expand our effective action along the lines of non relativistic GR (NRGR) \cite{Goldberger:2004jt} and each term in the action will scale as explicit powers of $v/c$ and $v_R/c_s$. This is in contrast with the formulation of \cite{Porto} where the spin terms in the effective action contain all powers of $v_R/c_s$.  However, this exposes a possible limitation of our approach. As written, our approach cannot be used to describe maximally rotating objects---their rotational frequency is precisely such that all the terms in the derivative expansion become of the same order. This is a symptom of expanding around the wrong background---a maximally rotating object being ``maximally away'' from an object at rest. In order to describe such black holes we would need to match coefficients at every order in the EFT and notice that we can perform a cumbersome resummation. Meanwhile, the formalism of \cite{Porto} can easily handle the maximally rotating case: precisely because the couplings of gravitons to the spin degrees of freedom contain contributions of all orders in $v_R/c$, which in this case is $\sim 1$. From this point of view, one can think of the action in \cite{Porto} as being a resummed version of ours. Hence, we believe that the two constructions lead to complementary results: the effective action outlined in this paper being appropriate for working with slowly rotating objects and that of \cite{Porto} being ideal for maximally rotating ones.

Another interesting point of comparison between the algebraic method employed in this paper and the explicitly covariant approach of \cite{Porto,Porto:2006bt,Porto:2008tb,Hanson:1974qy} is how the redundant degrees of freedom are eliminated. In the covariant approach, one imposes the constraint equation $S^{\mu \nu}p_{\nu}=0$, were $S$ and $p$ are the conjugate momenta associated with the rotational and translational degrees of freedom respectively. Such constraint has a clear physical interpretation, but its explicit form depends on the coefficients in the Lagrangian  \cite{Hanson:1974qy} and it must be solved anew at every order in the derivative expansion. In our algebraic approach, the redundant degrees of freedom are eliminated by imposing the inverse Higgs constraint $\nabla \pi^i=0$. It is insensitive to the details of the Lagrangian, can be solved once and for all and its solution is valid at all orders in perturbation theory. It would be important to understand whether it is possible to combine the advantages of both approaches and develop a theory with a Lagrangian-independent constraint and the ability to describe rapidly rotating objects. We leave this question for future study.

In Ref.~\cite{follow_up_paper} we will develop further the results in this paper and implement explicitly the NRGR expansion to describe objects that both move {\em and} rotate slowly. Additionally, we will discuss more fully the extension of our algebraic construction to include important higher order effects such as finite size and dissipative couplings \cite{Goldberger:2004jt, Porto:2006bt, GR_diss, Porto:2007qi, ENPW_diss}. And finally, we will derive explicit diagrammatic rules that can be used to systematically calculate observables to any desired order $v/c$ and $v_R/c_s$.



\section*{Acknowledgments}

We would like to thank Tomas Brauner, Garrett Goon, Walter Goldberger, Bart Horn, Lam Hui, Ian Low, Riccardo Rattazzi, Rachel Rosen and especially Alberto Nicolis for stimulating discussions. In addition, we are grateful to Tomas Brauner, Paolo Creminelli, Alberto Nicolis, Rachel Rosen and Ira Rothstein for valuable comments on an early draft of this manuscript. FR  acknowledges support from the Swiss National Science Foundation, under the Ambizione grant PZ00P2 136932. The work of RP was supported by NASA under contract NNX10AH14G and by the DOE under contract DE-FG02-11ER41743. The work of AM is supported by the Swiss National Science Foundation.



\appendix

\section{Notation}

Let us briefly summarize the major conventions and results that we will use throughout this paper. 
We will often have to distinguish between space-time and local Lorentz indices. Throughout this paper (with the notable exception of Section \ref{membranes}):
\begin{itemize}
\item $\mu, \nu, \sigma, \delta \ldots$ indicate (possibly curved) \emph{space-time} indices in  $d+1$ dimensions,
\item $a, b, c, d, \ldots$ indicate (flat) \emph{ Lorentz} indices in $d+1$ dimensions,
\item $i, j, k, l \ldots$ indicate \emph{spatial Lorentz} indices in $d+1$ dimensions.
\end{itemize}

We use a space-time metric with ``mostly plus'' signature, i.e. $\eta_{ a b}=\text{diag}(-,+,\ldots,+)$. The algebra of the Poincar\'e group is then given by:
\bea
\left[P_a, P_b \right]&=&0 \label{PPcommutator}\\
\left[P_a ,J_{b c}\right]&=& i (P_b \eta_{a c}-P_{c} \eta_{a b}) \label{PJcommutator}\\
\left[J_{a b},J_{c d}\right]&=& i \left[ \left(J_{ b d} \eta_{a c}-(a \leftrightarrow b)\right)-(c \leftrightarrow d) \right] \;. \label{JJcommutator}
\eea
In the fundamental (vector) representation, the generators for the Lorentz transformation are given by
\be
(J_{ab})_{cd}= - i\left(\eta_{ac} \eta_{bd}-\eta_{ad} \eta_{bc} \right), \;
\ee
and infinitesimal Lorentz transformations are therefore
\be
\label{inf Lorentz trans}
\Lambda^a \, _b =  ( e ^ {\frac i 2 \alpha ^ {cd} J_ {cd}} )^ a{}_b = ( e^ \alpha )^ a{}_b \approx \delta^a_b + \alpha^ a{}_b\; .
\ee

Throughout some of the calculations done in this paper the expression
\be
\text{Tr}\left[J^{ab} J^{ef} \right]=2(\eta^{ae}\eta^{bf}-\eta^{af}\eta^{be}) \;.
\ee
is quite useful. Additionally, we define the boost vector as
\be
K^i \equiv J^{i0} \; 
\ee
and in $3+1$ space-time dimensions we define the rotation vector as
\be
J_i \equiv \frac{1}{2} \epsilon_{ijk} J^{jk} \;.
\ee 

Acting on states, a unitary operator representing a Lorentz transformation $U(\Lambda) = e ^ {\frac{i}{2} \alpha J }\cong I+\frac{i}{2} \alpha ^ {ab} J _{ab}$ obeys:
\be
U(\Lambda \Lambda')=U(\Lambda)U(\Lambda')
\ee
 and thus if we examine $U(\Lambda)^{-1}U(\Lambda')U(\Lambda)=U(\Lambda^{-1}\Lambda'\Lambda)$ we conclude that 
\bea
\label{transformation of J}
U(\Lambda)^{-1}J^{a b}U(\Lambda)&=&\Lambda^a \, _c \Lambda^b \, _d J^{c d}\\
\label{transformation of P}
U(\Lambda)^{-1}P^{a}U(\Lambda)&=&\Lambda^a \, _b P^ b \;.
\eea

\section{Poincar\'e as an Internal Symmetry}\label{gravity_and_coset:app}

The usual geometrical description of gauge theories begins with the introduction of the principal bundle $P(M,G)$ with base manifold $M$ (space-time) and a structure group $G$. In the case of gravity we take the Poincar\'e group, $G=ISO(1,3)$. Matter fields are realized as sections of different associated fiber bundles. In this approach the action of the two symmetries of the system, namely the Poincar\'e group and diffeomorphisms is separated. The coordinates $x ^ \mu$ that describe the position on the base manifold $M$ only transform under diffeomorphisms, but have no action under the local Poincar\'e group. In other words the diffeomorphisms can be viewed as relabeling the points on the base manifold, while the local Poincar\'e transformation is a transformation along the fiber. Therefore, under infinitesimal diffeomorphisms $x ^ {\prime \mu} =  x ^ \mu + \xi ^ \mu (x) $, and we have
\begin{subequations}
\be
\phi' (x)   =   \phi (x) - i \xi ^ \mu (x)  \, \partial _ \mu \phi (x)\, .
\label{diffs}
\ee
\end{subequations}

To keep local $G$-invariance manifest, gauge fields corresponding to the Poincar\'e group are introduced as in \eq{SuperMaurerCartan} and the following transformation properties under the group $G$
\be
\ba{llrl}
g&=&e ^ {i a P}: 
&
\l \{
\ba{lll}
\tilde e ^ {'a} _ \mu & = & \tilde e ^ a _ \mu - \tilde \omega ^ {a} _ {\mu \, b} a ^ b - \partial _ \mu a ^ a\, , \\
\tilde \omega ^ {'ab} & = & \tilde \omega ^ {ab}\,, \\
\ea
\r.
\\
g&=&e ^ {\f {i} {2} \a J}: 
&
\l \{
\ba{lll}
\tilde e ^ {'a}  & = & \Lambda ^ a \, _ {b} \tilde e ^ b = \tilde e ^ a + \a ^ a _ {~b} \tilde e ^ {b}\, , \\
\tilde \omega ^ {'ab} _ \mu & = & \Lambda ^ a \, _ {c} \Lambda ^ b \, _d \, \tilde \omega ^ {c d} _ \mu 
+ ( \Lambda \partial _ \mu \Lambda ^ {-1} ) ^ {ab}
= \tilde {\omega} ^ {ab} _ \mu + \tilde \omega _ \mu ^ {ac} \a ^ b _ {~c}
+ \tilde \omega _ \mu ^ {cb} \a ^ a _ {~c} - \partial _ \mu \a ^ {ab} \, ,
\label{gauge_transform}
\ea
\r.
\ea
\ee
where indices are raised and lowered with the Minkowski metric $\eta_{ab}$. The fields appearing in \eq{SuperMaurerCartan}
are related by
\bea
e _ \mu ^ a & = & \tilde e _ \mu ^ a + \partial _ \mu y ^ a + \tilde \omega ^ a _ {\mu \, b} \, y ^ b \nn \\
 \omega_\mu^{ab}&=& \tilde \omega_\mu^{ab}\,.
 \label{eandomega}
\eea
Under local translations, both $e_\mu^{a}$ and $\omega_\mu^{ab}$ (defined as the coefficients of $P_a$ and $J _{ab}$ in \eq{SuperMaurerCartan}) are singlets, while under the local Lorentz group $e_\mu^{a}$ transforms linearly, and $\omega_\mu^{ab}$ as a connection:
\bea\label{trafoew}
e_\mu ^ {'a} & = &   \Lambda^ a _ {~b} e_\mu ^ b, \nn \\
\omega ^ {'ab} _ \mu & = & \Lambda ^ a _ {~c} \Lambda ^ a _ {~d}  \, \omega ^ {c d} _ \mu 
+ \Lambda ^ a _ {~c} \partial _ \mu ( \Lambda ^ {-1}) ^ {c b}.
\eea
At the same time, under the diffeomorphisms defined in \eq{diffs} the transformation of the $e ^ a _ \mu$ field reads
\be\label{dididiffs}
\delta ^ {\text{Diffs}} e _ \mu (x) = - e _ \n (x) \partial _ \mu \xi ^ \n (x) - \xi ^ \n (x) \partial _ \n e _ \mu (x),
\ee
and precisely the same for $\omega ^ {a b} _ \m$. Eqs.~(\ref{trafoew}) and (\ref{dididiffs}) coincide with the transformation properties of a vierbein  $e _ \m ^ a$ and a spin-connection $\omega ^ {a b} _ \m$, when the diffeomorphisms of \eq{diffs} are thought of as the translational part of the Poincar\'e group.

The fields $\tilde e _ \m ^ a$ and $\tilde \omega ^ {a b} _ \m$ are the necessary ingredients to describe a theory invariant under the local action of the Poincar\'e group. Considering the curvature tensor associated with $G$ we find
\bea
\label{G_curvature}
\mathcal R _ {\mu \nu} = \l [ {D}_ \m, {D} _ \n \r ] = i \tilde T ^ {a} _ {\m \n} P _ a + \f {i} {2} \tilde R ^ {a b} _ {\m \n} J _ {ab}
& = &
i \l ( \partial _ \m \tilde e ^ a _ \n - \partial _ \n \tilde e ^ a _ \m 
+ \tilde e _ {\m b} \tilde \omega ^ {a b} _ {\n} - \tilde e _ {\n b} \tilde \omega ^ {a b} _ {\m} \r ) P _ a \\
& + & 
\f {i} {2} \l ( \partial _ \mu \tilde \omega ^ {a b} _ \n - \partial _ \n \tilde \omega ^ {a b} _ \m
+ \tilde \omega _ {\m c} ^ a \tilde \omega ^ {c b} _ \n - \tilde \omega _ {\n c} ^ a \tilde \omega ^ {c b} _ \m \r ) J _ {a b}. \nn
\eea
Note that under the local shifts (\ref{gauge_transform}) $\tilde T _ {\m \n}$ and $\tilde R _ {\m \n}$  do not transform independently,
\bea
\tilde T _ {\m \n} ^ {'a} & = & \tilde T _ {\m \n} ^ {a} - \tilde R _ {\m \n} ^ {a b} \, a _ b, \nn \\
\tilde R _ {\m \n} ^ {' a b} & = & \tilde R _ {\m \n} ^ {a b}.
\eea
This suggests that we define new (gauge transformed) tensors
\be
\Omega ^ {-1} \l [ {D} _ \m, {D} _ \n \r ] \Omega = i ( \tilde T ^ {a} _ {\m \n} + \tilde R ^ {a b} _ {\m \n} \, y _ b) P _ a + \f {i} {2} \tilde R ^ {a b} _ {\m \n} J _ {ab} \equiv i T ^ {a} _ {\m \n} P _ a + \f {i} {2} R ^ {a b} _ {\m \n} J _ {ab}\,,
\label{pi_gauge}
\ee
with $\Omega$ defined in \eq{SuperMaurerCartan}. Now, by construction, $T _ {\m \n}$ and $R _ {\m \n}$ transform independently and we denote them Torsion and Curvature respectively. These are the nicely behaving tensors which can be used to build the Lagrangian (\ref{EH_gravity0}).

In the flat space-time limit $\mathcal R _ {\m \n} = 0$ and it is easy to show that
\be
\tilde e ^ a _ \mu = - \partial _ \m a ^ a \quad \text {and} \quad \tilde \omega ^ {a b} _ \mu = 0\, ,
\ee
and as a result the formula (\ref{eandomega}) reduces to
\be
e ^ a _ \mu = \partial _ \mu y ^ a.
\ee
It is clear that in this case, using diffeomorphisms, one can always choose coordinates $x^ \m$ such that
\be
e ^ {' a} _ \m (x') = e ^ a _ \nu (x) \f {\partial x ^ \nu} {\partial x ^{' \mu}} 
= \partial _ \nu y ^ a\f {\partial x ^ \nu} {\partial x ^{' \mu}} = \delta ^ a _ \mu.
\ee
In other words, $x ^ \m = y ^ a \delta ^ \m _ a$.

\section{Recovering Geometry for the Membrane} \label{geometry for membrane}

While the discussion in Section \ref{membranes} in certainly complete, it is interesting to point out the physical, or rather, geometrical meaning of the results that have simply fallen out of our algebraic construction. This is clearest in flat space, and the extension to curved space is straightforward.

First note that, thanks again to the inverse Higgs constraint,  the coset ``vierbein'' $ E_\mu{}^\alpha = \d_\mu Y^A  \Lambda_A{}^\alpha (\xi)$ behaves really like a geometric vierbein for the induced metric, i.e.
\be\la{brane ind metr 2}
h_{\mu\nu} = E_\mu{}^\alpha  E_\nu{}^\beta \eta_{\alpha\beta}.
\ee
The vierbein $E_\mu{}^\alpha$ is not the only geometric quantity that arises naturally from the coset construction. The constraint $ \d_\mu Y^A  \Lambda_A{}^d=0$ means that the $(d+1)$-vector $\Lambda_A{}^d (\xi) $ must be perpendicular to all the $ \d_\mu Y^A $s. Moreover, since $\Lambda$ is a Lorentz transformation we have $\Lambda_A{}^d \Lambda^A{}^d = \eta^{dd} = 1$, and thus $\Lambda_A{}^d$ is a \emph{unit} vector. For a codimension-1 brane, there is only one unit vector $n_B$ that is perpendicular to all the $\d_\mu Y^A$ (and therefore to the membrane), and it is given by $n_B \sim \epsilon_{A_1 ... A_d B} \epsilon^{\mu_1 ... \mu_d} \d_{\mu_1} Y^{A_1} ... \, \d_{\mu_d} Y^{A_d} = \delta_B^d - \delta_B^\mu \d_\mu \pi$. By requiring that this vector has unit norm, we thus get
\be
\Lambda_A{}^d (\xi) \equiv n_A = \fr{ \delta_A^d - \delta_A^\mu \d_\mu \pi}{\sqrt{1 + (\d \pi)^2}}.
\ee

From a geometric point of view, it is natural to consider how the direction of the normal unit vector $n_B$ varies from place to place on the membrane---or equivalently, how the membrane is embedded in the bulk. This information is encoded in the extrinsic curvature of the brane, which is defined as the covariant derivative of the normal vector ``projected'' on the brane:
\be \la{ext curv brane}
K_{\mu\nu} =  \d_\mu Y^A \d_\nu Y^B \nabla_A n_B = - \fr{\d_\mu \d_\nu \pi}{\sqrt{1 + (\d \pi)^2}}.
\ee
In the last step we used the fact that $\nabla_A = \d_A$ in the absence of gravity in the bulk. This quantity is clearly of higher order in the derivative expansion compared to the induced metric (\ref{brane ind metr}), and it is interesting to see how it arises from the coset construction.  To this end, let us consider the covariant object $\nabla_\alpha \xi_\beta$. After plugging in the solution to the inverse Higgs constraint $\nabla_\alpha \pi =0$, these covariant derivatives become higher order in the derivative expansion, and are thus a natural candidate to recover the extrinsic curvature. From equations (\ref{MC brane}), we get
\be \la{nabla xi 1}
\nabla_\alpha \xi_\beta = (E^{-1})_\alpha{}^\mu (\Lambda^{-1})_\beta{}^B \d_\mu n_B.
\ee
We can calculate $ (E^{-1})_\mu{}^\nu $ explicitly by using the relation
\be \la{brane E inverse}
E_\mu{}^\alpha (\Lambda^{-1})_\alpha{}^C = \d_\mu Y^A  \Lambda_A{}^\alpha (\Lambda^{-1})_\alpha{}^C  = \d_\mu Y^A  \Lambda_A{}^B (\Lambda^{-1})_B{}^C = \d_\mu Y^C,
\ee
where in the second step we used again the inverse Higgs constraint. For $C= \gamma$, this equation shows that $ (E^{-1})_\alpha{}^\mu = (\Lambda^{-1})_\alpha{}^\mu $, and since the normal vector $n_A$ depends only on the coordinates on the brane, we can rewrite equation (\ref{nabla xi 1}) in a more symmetric form:
\be  \la{nabla xi 2}
\nabla_\alpha \xi_\beta =  (\Lambda^{-1})_\alpha{}^A (\Lambda^{-1})_\beta{}^B \, \d_A n_B.
\ee
This almost looks like the definition of the extrinsic curvature (\ref{ext curv brane}), but not quite. The reason is that, according to the coset construction procedure, we can now build invariant quantities by contracting the covariant derivatives (\ref{nabla xi 2}) with the Minkowski metric $\eta_{\alpha\beta}$ on the brane. Instead, in the geometric picture scalar quantities are built by contracting the indices of the extrinsic curvature (\ref{ext curv brane}) using the induced metric $h_{\mu\nu}$ in (\ref{brane ind metr 2}). Equivalently, we can also use the rule of thumb that ``space-time'' indices $\mu, \nu, ...$ should be contracted with $h_{\mu\nu}$ whereas ``Lorentz'' indices $\alpha, \beta, ...$ should be contracted with $\eta_{\alpha\beta}$. In the end, because of equation (\ref{brane ind metr 2}) the difference is just a factor of $E_\mu{}^\alpha$ per index, and in fact we can use equation (\ref{brane E inverse}) to  get
\be
E_\mu{}^\alpha E_\nu{}^\beta \nabla_\alpha \xi_\beta = \d_\mu Y^A \d_\nu Y^B \d_A n_B = K_{\mu\nu}.
\ee

Finally, let us understand how to build the covariant derivatives for the matter fields living on the brane. Once again, according to the coset construction~\cite{ogievetsky:1974ab}, the covariant derivatives of matter field $\psi$ that transforms according to some (possibly reducible) representation of the Lorentz group on the brane is
\be \la{brane mat cov dev}
\nabla_\alpha \psi = (E^{-1})_\alpha{}^\mu \l(\d_\mu \psi + \fr{i}{2} (\Lambda^{-1})^\beta{}_C \, \d_\mu \Lambda^{C\gamma} J_{\beta\gamma} \psi\r).
\ee
Once again, the factor of $(E^{-1})_\alpha{}^\nu$ on the RHS is there because if we contract $\nabla_\alpha \psi$ with, say, $\nabla_\alpha \xi_\beta$ using the Minkowski metric, this should correspond to a contraction between a covariant derivative on the brane and the extrinsic curvature performed using the induced metric. 
From a geometric point of view, covariant derivatives of matter fields should be built using the connection induced on the brane. In other words, the factor $(\Lambda^{-1})^\beta{}_C \, \d_\nu \Lambda^{C\gamma}$  in equation (\ref{brane mat cov dev}) must be equal to the spin connection associated with the vierbein $E_\mu{}^\alpha$. This can be proven explicitly, although it requires a few manipulations. First, we can rewrite 
\bea
(\Lambda^{-1})^\beta{}_C \, \d_\nu \Lambda^{C\gamma} = (\Lambda^{-1})^\beta{}_C \, \d_\nu (\Lambda^{-1})^{\gamma C} 
&=& (E^{-1})^{\beta\mu} \d_\mu Y_C \, \d_\nu \l[ (E^{-1})^{\gamma\sigma} \d_\sigma Y^C\r] \la{spin brane step 1} \nonumber  \\
&=&  (E^{-1})^{\beta\mu} \l( h_{\mu\sigma} \d_\nu+ \d_\mu \pi \d_\nu \d_\sigma \pi \r) (E^{-1})^{\gamma\sigma}. \la{spin brane step 2}
\eea
where in the second step we used equation (\ref{brane E inverse}), and in the last step we used the definition of the induced metric. Now, according to~\cite{poisson2004relativist}, the Christoffel connection associated with the induced metric obeys the following equation:
\be
h_{\mu \rho}\Gamma^{\rho}_{\nu\sigma} = \d_\mu Y^C  \d_\sigma Y^B \nabla_B \d_\nu Y_C = \d_\mu \pi \d_\nu \d_\sigma \pi,
\ee
where in the last step we used again the explicit form of the induced metric, together with the fact tha $\nabla_A = \d_A$ and that $\pi$ depends only on the coordinates on the brane. Thus, we can rewrite (\ref{spin brane step 2}) as
\be
 (\Lambda^{-1})^\beta{}_C \, \d_\nu \Lambda^{C\gamma} = (E^{-1})^{\beta\mu}  h_{\mu\rho}\l( \delta^\rho_\sigma \d_\nu + \Gamma^\rho_{\nu \sigma} \r) (E^{-1})^{\gamma\sigma},
\ee
This is precisely the form of the spin connection given in~\cite{weinberg:1972bo}. Thus, the covariant derivative (\ref{brane mat cov dev}) is equivalent to the one that can be defined using the spin connection associated with the induced metric.


\small

\bibliographystyle{mine}
\bibliography{gravitycoset}


\end{document}